# Deployment corrections

## An incident response framework for frontier AI models

Institute for AI Policy and Strategy (IAPS)

**AUTHORS**
Joe O'Brien - Associate Researcher
Shaun Ee - Researcher
Zoe Williams - Research Manager

# Table of Contents

# Abstract

A comprehensive approach to addressing catastrophic risks from AI models should cover the full model lifecycle. This paper explores **contingency plans for cases where pre-deployment risk management falls short:** where either very dangerous models are deployed, or deployed models become very dangerous.

Informed by incident response practices from industries including cybersecurity, we describe a toolkit of deployment corrections that AI developers can use to respond to dangerous capabilities, behaviors, or use cases of AI models that develop or are detected after deployment. We also provide a framework for AI developers to prepare and implement this toolkit.

We conclude by recommending that frontier AI developers should (1) maintain control over model access, (2) establish or grow dedicated teams to design and maintain processes for deployment corrections, including incident response plans, and (3) establish these deployment corrections as allowable actions with downstream users. We also recommend frontier AI developers, standard-setting organizations, and regulators should collaborate to define a standardized industry-wide approach to the use of deployment corrections in incident response.

**Caveat:** This work applies to frontier AI models that are made available through interfaces (e.g., API) that provide the AI developer or another upstream party means of maintaining control over access (e.g., GPT-4 or Claude). It does not apply to management of catastrophic risk from open-source models (e.g., BLOOM or Llama-2), for which the restrictions we discuss are largely unenforceable.

# Executive Summary

To manage catastrophic risks[1] from frontier AI models[2] that either (a) slip through pre-deployment safety filters,[3] or (b) arise from improving the performance of deployed models,[4] **we recommend that leading AI developers establish the capacity for "deployment corrections"** in response to dangerous behavior, use, or outcomes from deployed models, or significant potential for such incidents.

We argue that **deployment corrections can be broken down into the following categories:**

1. **User-based restrictions** (such as blocklisting specific users or user groups);
2. **Access frequency limits** (such as limiting the number of outputs a model can produce per hour);
3. **Capability or feature restrictions** (such as filtering outputs or reducing a model's context window);
4. **Use case restrictions** (such as prohibiting high-stakes applications); and
5. **Model shutdown** (such as full market removal or the destruction of the model and associated components).

Frontier AI developers can mix and match these tools based on the threat model– for example, filtering outputs may be especially suited to preventing the spread of dangerous biological or chemical designs, while access frequency limits could be used to reduce the scale of some model-based incidents by limiting the rate of a model's outputs (for example, by reducing the speed of misinformation production). We envision deployment corrections as a toolbox that can be adjusted according to the type and severity of risks presented by each case.

---

[1] "Catastrophic risk" from AI models can be defined in several ways; Barrett et al. (2023) (p.22-23) includes the term in a tentative impact assessment scale for AI model development or deployment: "A severe or catastrophic adverse effect means that, for example, the threat event might: (i) cause a severe degradation in or loss of mission capability to an extent and duration that the organization is not able to perform one or more of its primary functions; (ii) result in major damage to organizational assets; (iii) result in major financial loss; or (iv) result in severe or catastrophic harm to individuals involving loss of life or serious life-threatening injuries."; according to Koessler & Schuett (2023), "By the term 'catastrophic risk' we loosely mean the risk of widespread and significant harm, such as several million fatalities or severe disruption to the social and political global order [...] This includes 'existential risks', i.e. the risk of human extinction or permanent civilizational collapse." For this paper, we follow the latter definition by Koessler and Schuett.

[2] Definition drawn from Anderljung et al. (2023): "highly capable foundation models for which there is good reason to believe could possess dangerous capabilities sufficient to pose severe risks to public safety [...] Any binding regulation of frontier AI, however, would require a much more precise definition."

[3] Such filters include, for example, staged release, alignment techniques, and model evaluation for extreme risks (Anthropic, 2022; Shevlane et al., 2023; Solaiman et al., 2019).

[4] Several pieces review methods that allow for improving the performance of deployed models: See Anderljung et al. (2023) (p.12) for an overview; Villalobos & Atkinson (2023) also reviews methods for improving an existing model's capabilities (at the cost of increasing inference compute use). Importantly, such discoveries can happen long after a model is initially deployed–meaning that systems warranting deployment corrections may be integrated into many downstream systems. Developers should therefore be careful to manage expectations, liability, and risk for downstream systems, especially in safety-critical use cases. See further discussion on this point in Sec. 3.1: Preparation, and Sec. 3.4: Recovery & follow-up.

**We then describe a high-level deployment correction framework for AI developers,** outlining a four-part process inspired by incident response practices from the cybersecurity field:[5] preparation, monitoring & analysis, execution, and recovery & follow-up.

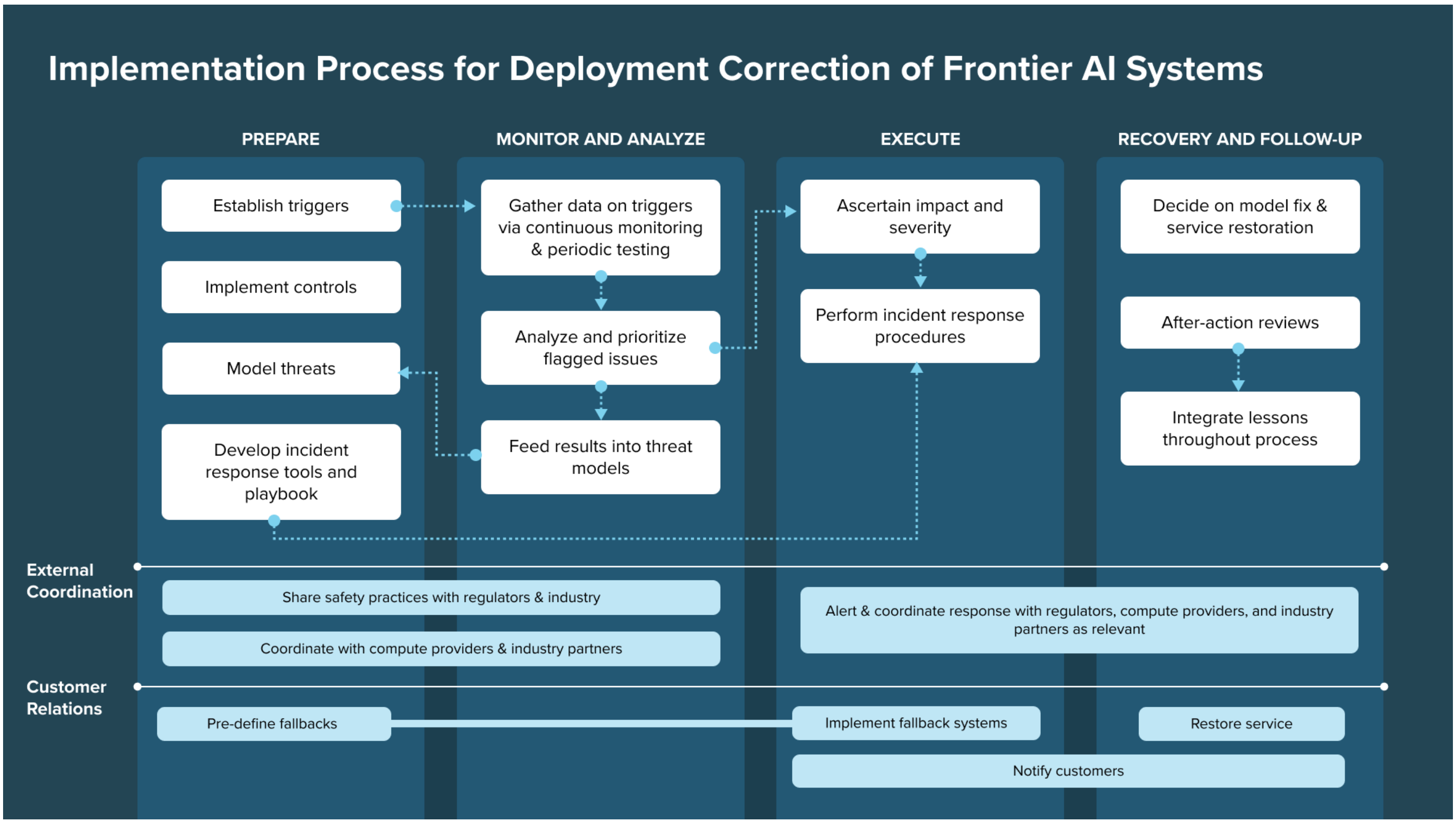

*Figure 1. An end-to-end process for implementing deployment corrections for frontier AI models*

1. **Preparation** refers to the act of building and adopting the tools and procedures that will allow an AI developer to act swiftly and effectively in response to an incident. It includes identifying and understanding possible threats, establishing triggers for deployment corrections, developing tools and procedures for incident response, and establishing decision-making authorities. Externally, it includes sharing insights on best practices with regulators and industry partners and defining fallback options for downstream users in the case of service interruption.
2. **Monitoring** refers to the process of continuously gathering data on a model's capabilities, behavior, and use (via a diverse range of sources), analyzing this data for anomalies, and escalating cases of concern to relevant decision makers. AI developers should also feed relevant data back into the threat modeling process.

[5] As a general note, relevant best practices have already been developed over years by organizations working in incident response and cybersecurity, such as the National Institute of Standards and Technology (NIST); we have noted throughout the document where specific guidance documents may be of use, and recommend that frontier AI developers and policymakers should draw on those resources when determining approaches to incident response for frontier AI.

3. **Execution** refers to the decision to apply a deployment correction to a model and the procedures that follow this decision. This stage also includes alerting and coordinating with relevant regulatory authorities, implementing fallback systems for downstream users, and notifying customers of the situation.
4. **Post-incident followup** refers to the set of actions relevant to recovery, restoration, learning, and ongoing risk management in the wake of an incident. This stage involves the process of repairing a model and restoring service, after-action reviews, and feeding lessons back into the previous stages. In some cases, this stage may require significant involvement from external parties (such as in the case that the incident is particularly severe *and* likely to occur in models developed by other companies).

**We then review several challenges to effectively using deployment corrections.** There are several unique challenges that frontier AI poses to this process.

- Challenges to threat identification include: (a) catastrophic risks from AI are complex and are marked by high uncertainty; (b) the landscape of frontier AI is rapidly changing; and (c) it is unclear how to assess deployed AI models for less acute risks.
- Challenges to monitoring include: (a) achieving monitoring coverage across the digital infrastructure may be a complex task, (b) frontier AI developers may face data overload, (c) designers of monitoring and alert systems must attempt to minimize both false positives and false negatives, and (d) advanced threat actors, and/or frontier AI models, may be able to evade standard monitoring mechanisms.
- Challenges to incident response include: (a) automated systems can fail rapidly, and (b) deployment corrections can only address issues if the model remains under control of the organization (currently, this is only the case for a subset of AI developers).
- Frontier AI developers may also face disincentives to deployment corrections, such as reputational, financial, and legal risks. Competitive pressures may add additional difficulty to following best practices. Without industry-wide collaboration, companies that prioritize safety (e.g., preemptively pull a model from market) may be outpaced by competitors who do not prioritize safety.

While we recommend some tentative ideas for mitigating these challenges, we believe these challenges will require significant work to solve, and encourage further work to do so.

**We close by recommending actions that frontier AI developers, policymakers, and other relevant actors can take** to lower the barrier for making decisive, appropriate deployment corrections, namely:

- Frontier AI developers should maintain control over model access, and carefully address situations in which partnering entities have access to model weights.

- Frontier AI developers should establish or grow dedicated teams to design and maintain processes for deployment corrections, including incident response plans and specific thresholds for response (we suggest Security Operations Centers[6] as a template, though the appropriate arrangement may vary company to company).
- Frontier AI developers should establish deployment corrections as allowable actions with downstream users, through the use of contracts and expectation-setting.
- Policymakers, standard-setting organizations, and frontier AI developers should establish a collaborative approach to managing deployment corrections. This could include information-sharing on threat models, developing secure communication channels, and managing incentives for effective incident response.

# Introduction

Some applied research on managing catastrophic risk from frontier AI models has focused on model risk assessments prior to public/commercial deployment (ARC Evals, 2023). However, there has been very little public work on *post-deployment* interventions. While several authors have discussed evaluation and monitoring for deployed models (Mökander et al., 2023; Shevlane et al., 2023), they have done so within broader discussions of model risk assessment and evaluation, and do not describe in depth the process for responding to situations in which *deployed* models fail evaluations or otherwise exhibit undesired behavior.[7]

The attention on pre-deployment risk assessment for frontier AI is warranted–modern best practices in engineering safety prioritize *designing out hazards*, rather than *responding to accidents* (Leveson, 2020). Still, because frontier AI poses potentially extreme impacts,[8] and because model capabilities and behaviors are hard to foresee even with pre-deployment testing, it requires a defense-in-depth approach (Ee, 2023). **To address this, this paper looks at contingency plans for cases where pre-deployment risk assessment falls short:** when either very dangerous models are deployed, or the continued availability of deployed models becomes very dangerous.[9]

[6] Security Operations Centers (SOCs) are dedicated security teams, typically running 24/7, tasked with a number of functions related to the security of an organization and its assets.
[7] Notably, the NIST AI RMF Playbook describes certain recommendations for AI developers who intend to "supersede, disengage, or deactivate AI systems that demonstrate performance or outcomes inconsistent with intended use" (NIST AIRC Team, n.d.). We believe the NIST playbook and associated resources will be useful.
[8] Such as spurring new pandemics or eroding society's ability to tell fact from fiction (Piper, 2023; Horvitz, 2022). For a far more extensive overview of catastrophic AI risks, see Hendrycks et al. (2023).
[9] Per Shevlane et al. (2023): "a model should be treated as highly dangerous if it has a capability profile that would be sufficient for extreme harm, *assuming* misuse and/or misalignment."

# 1. Challenge: Some catastrophic risks may emerge post-deployment

**Some potentially catastrophic risks may not be identified until after a model is deployed.** Recent history is full of cases where models have behaved or been used in unintended ways after model deployment (Labenz, 2023; Vincent, 2016; Heaven, 2022; Roose, 2023; Lanz, 2023). While pre-deployment red-teaming and risk assessment is likely to help, AI developers should anticipate that some issues will only be identified in the post-deployment phase (Shevlane et al., 2023). As models become more capable, such issues will present more significant risks.[10] We envision two main sources of post-deployment risk:

(a) Risks that are not identified in pre-deployment risk assessments
(b) Risks arising from improving the performance of deployed models

**On (a): Pre-deployment model risk assessment is unlikely to identify *all* catastrophic risks**, for several reasons. First, model risk assessment tools are in the early stages and will take time to develop; additionally, the broad space of applications for frontier AI models poses a significant challenge to assessing all potential significant risks.[11] Second, certain risks may only exist in a less-bounded context than pre-deployment testing, such as adverse interactions with other systems, models, or organizations, unexpected forms of misuse from malicious actors, and adversarial attacks on systems integrated with critical infrastructure. Third, power-seeking and/or deceptive AI might successfully infer the existence of evaluation or monitoring environments, and "play along" until it can successfully evade such filters (Hendrycks et al., 2023, p. 41). Fourth, adverse or risky outcomes may take time to develop (e.g., increased vulnerability due to job loss in critical industries, or the development of new methods of misuse). It also isn't clear that risk assessment will focus on systemic risks of widespread AI adoption, in addition to more acute risks.[12]

**On (b): Pre-deployment model risk assessment may be infeasible for assessing risks emerging from improving the performance of deployed models.** Major ways this could happen include:

---

[10] For additional concrete examples, one could look to some of the risks invoked in the recent White House AI lab commitments announcement: "Bio, chemical, and radiological risks, such as the ways in which systems can lower barriers to entry for weapons development, design, acquisition, or use; Cyber capabilities, such as the ways in which systems can aid vulnerability discovery, exploitation, or operational use, bearing in mind that such capabilities could also have useful defensive applications and might be appropriate to include in a system; [...] The capacity for models to make copies of themselves or 'self-replicate'" (The White House, 2023).
[11] Furthermore, it is also not guaranteed that such tools will be robustly designed and reliably used.
[12] While we believe that developers and/or external watchdogs should monitor for such effects, identifying techniques for this is outside the scope of this piece. We recommend Solaiman et al. (2023), which begins to lay out an approach to evaluating generative AI models for social impact.

- The AI research community may identify new methods for building on base models' capabilities, such as external tool use, new frameworks for agency,[13] or automated prompt engineering (Schick et al., 2023; Zhou et al., 2023).
- The ongoing release of incremental model updates may also present risk if such updates are not subject to pre-deployment risk assessment.

These updates and capability extensions could occur rapidly, relative to the months-long process for developing base models. They may be hard to predict and not fully accounted for in pre-deployment risk assessments.

The following scenario gives one example of how a potentially catastrophic risk could pass through pre-deployment checks, and how the framework we'll discuss could be applied to mitigate the negative outcomes.

### Case 1: Partial restrictions in response to user-discovered performance boost and misuse

*Includes: improving performance of deployed models; misuse; reversion to allowlisting; restricting access quantity.*

- **Context:** After a long period of safe commercial use, and in response to feedback from users, Company A decides to significantly raise Model A's number of prompts/hour available through their API. Users who have been experimenting with auto-GPT-style architectures find that loosening this restriction makes these tools finally "usable," by resolving the issue that these systems would be forced to stop several minutes into each hour. An explosion in innovation with the architecture occurs, with startups developing plug-and-play Auto-GPTs for mainstream customers, and the technology sees widespread adoption.

- **Incident:** Investigative journalists break a case finding opposition forces have been leveraging Model-A-powered agents to run a powerful destabilization campaign against a small nation's government. Separately, it becomes apparent that, although it's not clear who's been prompting them, a global network of Model-A-powered agents have been collaborating to uncover trade secrets for US computer chips through a combination of cyberattacks, spear phishing, and data analysis.

[13] See Weng (2023) for a description of how LLM-centered agents can be designed by decomposing "agency" into separate components, such as planning, memory, and tool use.

- **Detection:** The relevant team within Company A (e.g., a SOC) tracks this information, using public journalism as a data source.

- **Incident response:**
    - The SOC manager decides to escalate, bringing the matter to Company A's CISO and arranging an emergency meeting. There is an existing playbook for this scenario (i.e., a scenario in which users stretch the capabilities of a deployed model in a way that introduces novel, dangerous use cases).
    - The SOC analyzes relevant data and decides upon limitations that would address the issue, namely, reintroducing limitations on prompts/hour for Model A.
    - The prompts/hour restriction is initiated for all cases except for pre-allowlisted, safety-critical customers (e.g., commercial customers that use Model A in narrow contexts, such as emergency services or the energy sector).
    - Communications:
        - Customers are alerted to the restriction via email and the API portal, and safety-critical customers who were not pre-allowlisted are contacted to discuss fallback to other systems.
        - Relevant agencies and industry partners are informed (e.g., CISA, the Frontier Model Forum, and any AI-specific regulators that have been formed).

- **Followup:**
    - Threat actors are identified, and more restrictions and tracking are put in place to prevent a repeat occurrence of this or similar incidents. This takes several months, after which the number of prompts/hour restriction is lifted.

# 2. Proposed intervention: Deployment corrections

To manage the above risks, **we recommend frontier AI developers establish the capacity to rapidly restrict access to a deployed model, for all or part of its functionality and/or users.** This would facilitate appropriate and fast responses to a) dangerous capabilities or behaviors identified in post-deployment risk assessment and monitoring, and b) serious incidents.[14] We also recommend practices that can lower the barrier for making decisive, appropriate access restriction decisions–see the recommendations in Section 5.

The current section lays out access restriction options which allow for granular and scalable targeting based on the threat model (Section 2.1), and discusses additional considerations regarding cases of emergency shutdown (Section 2.2).

## 2.1 Range of deployment corrections

**Frontier AI developers which make their models available to downstream users via an API have a number of tools at their disposal to limit access to the model.** At a high level, this toolkit includes user-based restrictions, access frequency restrictions, capability restrictions, use case restrictions, and full shutdown. These tools can be used in a broad range of scenarios, from cases in which risks from the model are fairly limited,[15] to scenarios in which the harms are potentially severe and can arise even from proper use by an authorized (allowlisted) user.[16]

As discussed in Section 4, restricting model access may be difficult in practice, as downstream users may become dependent on capabilities of newly-deployed models.[17] To minimize these harms, and to lower the barrier for developers to institute deployment corrections as a precaution, we outline a space of deployment corrections to allow a *scalable* and *targeted* approach. AI developers can opt for combinations of user-based or capability-based restrictions, and tailor these choices to respond effectively to specific incidents, while minimizing downstream harms.

[14] It is worth placing this piece in the context of the recent Senate hearing on "Principles for AI Regulation," in which Stuart Russell (UC Berkeley), Dario Amodei (Anthropic), and Senator Richard Blumenthal discussed the necessity of developing and enforcing mechanisms to recall dangerous AI models from the market (*Oversight of A.I.*, 2023).
[15] For example: banning individual problem users; or in cases of embarrassing (but not catastrophic) model failures.
[16] For example: if the new model turns out to have reliability/security issues in critical infrastructure; the model has dangerous interactions with other autonomous agents or platforms; or if the model's capability is augmented in a relevant domain.
[17] The dependency problem will worsen over time as models are (a) adopted by more users, and (b) adopted in more sensitive use cases. In such cases, AI developers may face stronger disincentives from customers, shareholders, and possibly from regulators to impose deployment corrections on their models.

While we expect some staff at AI companies are familiar with these tools, we review them here because they will be referenced throughout this piece. The following table draws heavily on Shevlane et al. (2023)–particularly the appendix on deployment safety controls.

| Table 1: Taxonomy of deployment corrections | | |
|---|---|---|
| **#** | **Access Restriction** | **Description** |
| *1* | *User-based restrictions*[18] | |
| 1a | Blocklisting individuals or groups | Imposing IP or other verification-based restrictions on users based on anticipated or historical misuse. |
| 1b | Allowlisting individuals or groups | The inverse of blocklisting. Providing specific users or user groups expanded forms of access; this can be imposed at the time of deployment, or be imposed retroactively.[19] Maintaining an allowlist opens up the option to retain access for allowlisted users even when removing access for all others (e.g., due to widespread or unknown threat actors). |
| *2* | *Access frequency limits*[20] | |
| 2a | Throttle number of calls | Place a hard cap on the number of function calls (e.g., JSON documents sent to an external API) that a single model can output in a given amount of time. |
| 2b | Throttle number of prompts | Place a hard cap on the number of prompts that can be submitted to a model in a given amount of time. |
| 2c | Throttle number of end users | Place a hard cap on the total number of end users a model can have. |
| 2d | Throttle number of applications | Place a hard cap on the total number of applications that can be built on top of a model. |

[18] There is a question of how 'individuals or groups' are identified. Paid users are easier to identify and group, while second-order users (i.e., users of downstream applications) might be harder to identify, and require Know-Your-Customer and data-sharing policies.

[19] For an example of access restrictions designed into the deployment process, see Solaiman et al. (2019), or see OpenAI (2022) for an example of the use of private betas and use case pilots.

[20] Restrictions within this category may be imposed with a range of parameters, such as time spans (per day, per hour, etc.), user limits (e.g., number of prompts per user per hour), etc.

| | | |
|---|---|---|
| *3* | *Capability or feature restrictions* | |
| 3a | Reduce context windows | Reduce the number of tokens a model is capable of processing in relation to one another. This curbs a model's capabilities by reducing its ability to "remember" earlier information (Stern, 2023). |
| 3b | Session resets | Reset sessions after a certain number of prompts or outputs. This might accomplish a similar goal to the above point. |
| 3c | Limit user ability to fine-tune | "Fine-tuning," or re-training a base model to perform better at a particular task, might increase a model's capabilities in certain domains to the extent that such capabilities are dangerous. Frontier AI developers could remove this functionality for users, or retract specific fine-tuned instances. |
| 3d | Output filtering | Monitor and automatically filter out dangerous outputs, such as code that appears to be malware, or viral genome sequences. |
| 3e | Removal of dangerous capabilities | Attempt to remove specific capabilities (e.g., pathogen design) via fine-tuning, reinforcement learning from human feedback (Lowe & Leike, 2022),[21] concept erasure (Belrose et al., 2023), or other methods. |
| 3f | Global planning limits | Adjust whether the same model instance has access to a large number of users, or is limited to more narrow sets of interactions (Shevlane et al., 2023). |
| 3g | Autonomy limits | For example, restricting the ability for a model to define new actions (e.g., via assigning itself new sub-goals in an iterative loop), or to execute tasks (versus solely responding to queries) (Shevlane et al., 2023). |
| *4* | *Use case restrictions* | |
| 4a | Prohibiting high-stakes applications | Setting a use policy that restricts the model from being used in high-stakes applications, and allows banning or otherwise penalizing |

[21] It is worth noting that RLHF does not in fact directly remove dangerous capabilities, but instead can be used to *steer models away* from dangerous outputs. To the extent that this and other techniques *effectively* remove a model capability for downstream users, it may be reasonable to group such techniques in this category.

| | | |
|---|---|---|
| | | users that breach this policy.[22] Requires Know-Your-Customer procedures. |
| 4b | "Narrowing" a model | Producing fine-tuned/application-specific narrower models to reduce a model's capacity for general-purpose use. |
| 4c | Tool use limits | Limit the ability of a model to interact with downstream tools (e.g., to use other APIs), make function calls, browse the web, etc. |
| *5* | *Shutdown* | |
| 5a | Full market removal | Pull the current model from the market. Can also include pulling one or more previous versions, in cases where it is unclear whether reverting to a previous model would solve the issue. |
| 5b | Powering off | Disconnecting power to the relevant parts of the data center or cluster where the model is hosted. |
| 5c | Decommissioning | Decommission the model, including destroying data, systems, or assets associated with the model, whether through deletion of data or physical destruction.[23] |
| 5d | Moratorium | Institute a moratorium on re-deployment until approval via independent review. |

**The above options are not mutually exclusive–instead, they can be viewed as a toolbox that developers can mix and match** to address different threat models or incidents. For example, an AI lab might [1b] allowlist certain users (e.g., external auditors) for [3c] the ability to fine-tune a model and [4b] full model generality, while allowing other users access to the model but without those two capabilities. A developer may also wish to establish [3g] autonomy limits just in [4a] high-stakes applications.

**These options can be imposed manually, or triggered automatically.** It may make sense for certain restrictions to trigger automatically, such as in cases where the speed of failure is rapid, or to ensure that restrictions are imposed reliably in accordance with thresholds set forth in standards or pre-commitments. Similarly, manual triggers may be appropriate to ensure that human operators can act in cases where monitoring, automated response, or pre-determined thresholds fail to identify issues, or where careful deliberation is required. See Section 3 for further discussion on how the balance of manual and automatic decision-making can be managed.

[22] I.e., applications where the failure or removal of the model could result in significant harm (for example, self-driving cars).

[23] For decommissioning, developers might also turn to sources like the M3 Playbook Sec. 2.8: Develop a Decommission Plan or the CIO Decommissioning Template (though resources on specifically decommissioning *AI models* are scarce).

## 2.2 Additional considerations on emergency shutdown

**Emergency shutdowns** are common in areas where continued operation can result in catastrophic harm, such as nuclear energy (*Operating Reactor Scram Trending*, 2021), finance (*Circuit Breaker*, n.d.), and even in elevators (Palmer, 2023). The purpose is typically to intervene quickly to prevent an existing failure from resulting in a catastrophic outcome, by shutting down the affected system completely.

**In the case of frontier AI, companies may want to shut down models for a broad range of reasons**–some cases may be due to more obviously-dangerous issues, such as certain model-originating risks (e.g., deception or power-seeking), catastrophic forms of misuse, or severe social or economic effects; however, it is possible that an AI company might want to shut down models in cases of sub-catastrophic harm as well.[24, 25]

While developing fallbacks may mitigate some downstream harm, **shutdown is more likely than targeted restrictions to have severe repercussions for downstream users, up to and including breaking their applications (and leading them to switch over to the company's competitors).** In certain industries, these impacts may lead to loss of life or significant economic harms. Due to the potential scale of downsides for users, the reputational and financial costs to the AI developer, and the risk that safety-conscious companies will fall behind less safety-conscious competitors, additional support structures may be needed to incentivize appropriate risk management practices around shutdown. These could include regulatory oversight, industry standards, and/or financial incentives.

The following scenario, which features a temporary model shutdown, describes how AI developers might weigh this option against other deployment corrections.

---

[24] For example, one can look at existing cases of model shutdown, such as Microsoft's Tay (shut down due to toxicity), or Meta's Galactica (shut down due to hallucinations). While these cases illustrate that shutdown is not an *uncommon* response to AI model malfunction, one concern we have is that companies may become less willing to pull their models when such models are more deeply integrated into a broad set of downstream applications (for reasons discussed below). For contrast, Tay and Galactica were both pulled in 16 hours and three days, respectively, and so had not accumulated significant downstream dependencies.

[25] As discussed in the NIST AI RMF 1.0, organizations should define "reasonable" risk tolerances in areas where established guidelines do not exist; such tolerances might inform where the bar for shutdown should be. However, work in this area is nascent, especially for frontier AI models.

### Case 2: Full market removal due to improved prompt injection techniques

*Includes: Prompt injection; input-output monitoring; multi-agent interactions; full market removal*

- **Context:** While open-source models lag behind frontier models, they're widely-used and still highly competent. Last week, the weights and architecture of Company B's most powerful model yet ("OpenModel") were leaked, and it is now effectively open-source.

- **Incident:** Six months ago, Company C made Model C available via API. The base model is generally more capable than any existing model across a broad range of metrics, including scientific research capabilities. Company C has set careful input-output monitoring and filters on the model, and so far they've been successful in preventing misuse. However, users of OpenModel are able to use it to develop advanced prompt injection attacks to overcome Company C's filters. While automated output filters catch some of the attacks, Company C can't patch the holes fast enough. Reporting shows that Model C is outputting dangerous information of all kinds, the most worrisome including a process for developing airborne rabies.

- **Detection:** Company C's SOC's automated monitoring flags a notable uptick in prompt injection attacks and insufficiently-disguised dangerous model outputs; Twitter users near-simultaneously report successful stories of users circumventing input- and output-monitoring through clever use of prompt injection and encryption tools to elicit dangerous model outputs.

- **Incident response:**
    - Automatically, users flagged for prompt injection are booted from the platform.
    - Upon receiving reports of cases of many undetected prompt injections and dangerous output, SOC escalates to CISO and schedules an emergency meeting.
    - Company C decides to immediately remove all access to Model C. They consider retaining access for pre-allowlisted customers and for red-teamers, but conclude that the cyber capabilities of OpenModel could allow OpenModel users to hack into allowlisted accounts and continue to send prompt injection attacks from there.
    - Communications:

- Removal of Model C announced via all channels, and safety-critical customers are prioritized for triggering fallback to less capable systems and/or human operators.
        - Relevant agencies, industry partners, and regulators are informed. Emergency meetings are called to discuss if other models on the market need to be removed or rolled back to lower capability versions, given existing protections may be circumvented with OpenModel.

- **Followup**
    - Company C obtains a copy of OpenModel and uses it to adversarially train automated detection and response systems.
    - Company C also integrates tracking of new open-source AI models into the process of security maintenance, **establishing faster turnaround times for identifying and removing cyber threat of such models.**
    - Company C performs testing on the model to verify the issue has been addressed, and may work with external actors to certify these results.
    - Company C restores service to the model after taking the above steps.
    - Cybersecurity standards for developers of models that are as or more capable at cyberattacks than OpenModel are made more stringent.

# 3. Deployment correction framework

This section reviews implementation procedures for deployment corrections, drawing on tools and best practices from other industries as appropriate. We believe that staff at frontier AI companies will be familiar with much of the following, but that it is nevertheless valuable for us to describe in detail what is required for deployment corrections to function.

We will frame deployment correction as a four-part process, consisting of preparation, monitoring & analysis, incident response, and post-incident recovery & follow-up.[26]

- **Preparation** should prepare the organization and other relevant parties for the potential occurrence of a catastrophic risk. It will involve the organization proactively modeling threats, implementing controls to prevent (or mitigate the severity of) incidents, establishing triggers for deployment corrections, developing tools, and preparing an incident response plan with clearly defined roles and responsibilities. This allows the organization to act quickly and decisively when issues do arise.

[26] This process is inspired by the NIST computer security incident handling guide (Cichonski et al., 2012).

- **Monitoring & analysis** should aim to identify the occurrence of incidents or new risks as quickly as possible. It should involve regular testing of deployed models and gathering real-time data relevant to catastrophic risks identified in the preparation phase, tracking and prioritizing of incidents or cases of misuse, and feeding data back into both automated and manual assessment processes.
- **Execution** should aim to mitigate threats efficiently and fully. It will involve alerting key stakeholders, ascertaining the severity of the incident, performing deployment correction procedures to contain, remediate, and eliminate risks, and implementing fallback systems where necessary.
- **Recovery & follow-up** should aim to return systems to a safe state, and integrate lessons throughout the organization. It will involve a process for safely restoring service (depending on the severity and fixability of the incident), alerting external parties, notifying and providing remedy to customers, and running post-incident review to fix blind spots, including root cause analysis.

The process may also involve coordination and information-sharing with governments and industry partners (where such activities are likely to support effective incident response and do not violate relevant laws).

**Figure 1** (provided in the Executive Summary) provides an overview of this section–a tentative blueprint for the process that AI developers can adopt to integrate deployment corrections into their deployment process.

In the remainder of this section, we describe practices that will help AI developers to navigate each stage in the deployment correction process.

## 3.0 Managing this process

The process of incident response is complex, and will require the involvement of actors throughout the AI company (including product teams, business operations, safety engineers, and C-suite), as well as external parties, such as third-party auditors, other frontier AI developers, and government agencies. To improve coordination and allow for decisive action, we recommend centralizing the process under a clear owner.

**Security Operations Centers ("SOCs") may be an appropriate institutional home for much of this work.** Considering the complexity surrounding the deployment correction process, and the rapid rate of change in the field of AI, we believe that unifying security operations under one roof is sensible.[27]

SOCs at frontier AI companies should include some prominent functions from large SOCs in cybersecurity, including:

- **Analysis and monitoring of logs**. Logging instrumentation typically produces a large amount of data, the processing of which requires both automated tools and manual tools. This requires close collaboration between the SOC and application developers/users. While automation can support this function, human judgment and context are required. SOC analysts must work with developers and users who are more familiar with the actual application to calibrate the alerts and ensure they strike the right balance between minimizing false alerts and ensuring sufficient detection power.
- **Gathering and sharing threat intelligence**. Unlike monitoring, which in cybersecurity involves looking *within* a company's systems for signs of an intrusion (e.g., indicators of compromise), threat intelligence provides information on the observed behavior of threat actors; e.g., common techniques, tactics, and procedures (TTPs) that threat actor groups are using, or their current targets of interest.[28] For frontier AI developers, such "threat intelligence" could include real-world observations about TTPs to bypass model safeguards, ongoing campaigns by malicious actors that involve abuse of frontier AI models, or indicators of dangerous behavior by AI systems. Threat intelligence typically is provided by outside sources such as security vendors, community organizations, and governments.
- **Incident response**. See Computer Security Incident Response Teams (CSIRTs), and/or Computer Emergency Response Teams (CERTs) (Cichonski et al., 2012). These are the staff who respond "on the ground," and might include staff experienced in technical skills, such as analyzing malware or tracking the source of questionable behaviors from AI models.

**Note:** Throughout this section, we largely use the term "AI developers" rather than "Security Operations Centers" to refer to the acting entity, to leave to the discretion of specific developers *who* in their organization is assigned responsibility over which tasks; nevertheless, an SOC may be a reasonable owner for many functions related to mitigating risks from frontier AI models.

[27] It's possible that these tasks might not be housed in an SOC per se; for example, Trust & Safety teams may be positioned to tackle large parts of this process. Nevertheless, AI developers should be able to answer who within their company is responsible for these tasks and capable of handling them.

[28] Organizations with more mature cybersecurity practices may also engage in "threat hunting," which typically involves a specialized team using threat intelligence and other resources to proactively search for signs of an intrusion.

## 3.1 Preparation

Here, we describe in more detail how AI developers can make and maintain documented response plans for effectively using the toolbox of options for deployment corrections as outlined in Section 2. Policymakers and/or standard-setting organizations may also have a role in mandating or setting standards for AI developers to prepare tools and protocols for deployment correction, in order to overcome initial inertia and to incentivize adoption across the frontier AI industry.

**The preparation stage should involve:**

- Threat modeling;
- Instituting controls to prevent or mitigate the severity of incidents, including defining fall-backs for downstream users, especially in safety-critical domains;
- Establishing triggers for deployment corrections based on thresholds set and maintained as part of the threat modeling process;
- Developing a documented response plan for executing deployment corrections which clearly delineates roles and responsibilities;
- Ensuring that industry partners (such as partnering tech companies, and compute providers) adopt the above tools and protocols.

**AI developers should model potential catastrophic threats, and regularly update these threat models.** Threat models should trace high-level catastrophic risks to specific vulnerabilities (such as insider threats, poor handling of AI models, and cybersecurity vulnerabilities [e.g., authorization bypass]), and include mitigations for these vulnerabilities. To aid this process, AI developers may want to consider employing a set of risk assessment techniques from other industries (Koessler & Schuett, 2023). They may also want to involve external domain experts in the process of identifying specific threat models, such as was done with Anthropic's recent work on "frontier threats red teaming," which focused on biological risk (Anthropic, 2023). Risk identification, analysis, and evaluation are high-priority steps for risk management,[29] and it is important that frontier AI companies adopt a defense-in-depth approach that employs multiple overlapping techniques (Ee, 2023).

**AI developers and policymakers should develop a system of controls to prevent or mitigate the severity of incidents.** Other authors have explored a number of these controls extensively, such as pre-deployment risk assessment, third-party auditing, and AI alignment techniques.[30] We would like to make an addition to this list which pertains to the post-deployment phase: **AI developers should work with downstream applications and users to define fallback options, primarily in safety-critical use**

[29] For example, Barrett et al. (2023) lists several high-priority measures relating to risk assessment under Section 2.3 "High Priority Risk Management Steps and Profile Guidance Sections," such as "Identify whether a GPAIS could lead to significant, severe or catastrophic impacts" (guidance associated with Map 5.1 of the NIST AI RMF), or "Use red teams and adversarial testing as part of extensive interaction with GPAIS to identify dangerous capabilities, vulnerabilities or other emergent properties of such systems" (guidance associated with Measure 1.1 of the NIST AI RMF).

[30] See (Schuett et al., 2023) for an overview of these and other best practices.

**cases.**[31] In certain use cases, interrupted service could result in significant harm to downstream users. In such cases, developers and downstream users should work together to develop backup systems in the case of severe outages. Responses for safety-critical customers could involve a shift to allowlisting, rollback to a previous model, and/or fallback to non-AI software systems or even human operators.[32] Language requiring such plans to be in place should be written into terms of use or contractual agreements; a deployment correction clause could cover AI developers when they implement such corrections.

**AI developers should establish thresholds for initiating deployment corrections**, informed by the threat modeling process. Thresholds might be set by the AI developer and/or industry standards. An example case:

- Threat model: making biological weapon design easier and doable by more people.
- Thresholds: demonstrable evidence of someone using the AI model to design a novel dangerous pathogen, or using the model to design a benign biological agent via accessing similar capabilities as those that would be used in pathogen design.[33]
- Action: Emergency meeting is called; decision to switch AI model access to an allowlist of only safety-critical users, and for all other users to revert to a last-gen AI model until the exploit is resolved and/or the capability selectively removed.

**AI developers should create and maintain a documented incident response plan to guide the incident response process.** This document should clearly define the following aspects:

- Risk scenarios that warrant deployment corrections, as developed in the threat modeling process, and triggers to identify deviations from expected behavior.
- The composition of the response team, comprising representatives from IT, cybersecurity, AI development, legal, communications, relevant business units, and external domain experts. Due to the variety of potential risk scenarios, incident response may require expertise beyond what AI developers can handle alone, and require inputs from multiple parties.[34]
- The roles and responsibilities of different teams and individuals involved in the incident response process (in order to minimize chaos when handling an incident). To act swiftly, everyone on the team needs to know their responsibilities and the decisions that are theirs to make.

---

[31] See (*Governing AI: A Blueprint for the Future*, 2023) for more discussion on this point.
[32] However, it is worth noting that the fallbacks approach could, in some areas, be riskier and less advisable than limiting AI model involvement in the first place. For example, this may be true in the case of deciding whether to launch a nuclear weapon (*Buck, Beyer, Markey, and Lieu Introduce Bipartisan Legislation to Prevent AI From Launching a Nuclear Weapon*, 2023).
[33] For any given threat model, there may need to be multiple thresholds; for example, this threat model might also include thresholds around AI model capability in certain relevant domains (such as protein folding or virology).
[34] Additionally, AI developers should ensure that the response plan takes into account additional parties that may have access to the model, or otherwise have leverage over how the model is used–and potentially develop tools and protocols with these parties where appropriate. Relevant parties may include partnering tech companies that have access to model weights, and providers of computational resources used for model inference. The latter may have unique leverage over some aspects of monitoring and shutdown; for more on this, see Appendix I.

The response plan should be circulated to the developers or teams that handle the triggers for risks, and operators should be trained using these protocols to respond to a set of high-likelihood and/or high-consequence events.[35] This training should include procedures for risks or unusual behaviors that have not yet been identified, covering factors such as generic thresholds for severity, temporary mitigations that can be used during investigation, and appropriate escalation points. The response plan should also be updated periodically to incorporate changes in teams, personnel, AI technology, deployment correction tools, and risk models.

While we expect that Trust & Safety teams[36] at top AI companies will have experience in maintaining part of this suite of tools (as such companies already have some infrastructure for certain deployment corrections, as demonstrated by past actions[37] and documentation[38]), we are not aware that such tools are sufficient for a range of potentially catastrophic scenarios. This technical work is outside the scope of this piece.

**The response plan should explicitly define decision-making authority for deployment corrections**, with the design goal of ensuring that these actions are executed when needed but otherwise do not happen.[39] Recommending *how* authority should be divided is out of scope for this piece; however, we recommend that developers consider the following:

- **The extent to which decisions are automated, versus left to human operators**. Automated deployment corrections may be appropriate in certain cases–particularly where human intervention would be too slow. For example, a safety filter system should be authorized to automatically prevent an AI model from outputting text that explains how to design a novel pathogen–because by the time the text has been sent to a downstream user, the (potential) damage has been done. Exfiltration of model weights would be a similarly irreversible act. Where threat models and thresholds are clearly defined, AI developers might consider automating responses.[40] However, anticipating that risk assessment and management may include gaps due to the rapid pace of AI development and the large space of potential failures from increasingly general AI models, incident response will also need to rely on human judgment. AI developers should establish processes for manual triggering of deployment corrections as well, for cases where action may be required despite pre-specified thresholds not being met.[41]

[35] ISO/IEC 27035 and NIST Special Publication 800-61 Revision 2 provide additional guidance on incident response, and emphasize the importance of planning and training, among other supporting factors.
[36] Trust & Safety (T&S) teams typically work to maintain safe user experiences, often by addressing issues including privacy, bias, misuse, and harmful or illegal content, among other issues.
[37] For example, see OpenAI's geoblocking of Italy, blocklisting IP addresses, or taking its AI text classifier offline due to low accuracy.
[38] For example, see Anthropic's trust portal, or OpenAI's security portal.
[39] A helpful resource here may be found in Schuett (2022)–The author suggests a framework that AI developers can use to assign risk management roles and responsibilities, focusing on assigning responsibilities across product teams, risk & compliance teams, internal and external assurance parties, and at the board level.
[40] Examples from other industries where system failure could rapidly lead to catastrophic results include Reactor Protection Systems for nuclear power plants, which involve an intricate network of sensors and protocols designed to monitor for abnormal reactor signals and automatically trigger safe shutdown procedures as quickly as possible (USNRC HRTD, 2020), and failsafe systems for elevators, which trigger automatically in the case of loss of power (Palmer, 2023).

- **The extent to which authority is shared**. In certain industries, shutdown or other major operational changes can be triggered by any of multiple parties in response to safety concerns. One example of distributing shutdown authority is the concept of the Andon cord–a tool initially developed for the Toyota production line that allows any operator along the line to trigger a production pause (Tarlengco, 2023). Amazon has demonstrated adoption of this tool in digital spaces, by integrating Andon systems into their customer service (where Support agents have the authority to flag or pull a product from distribution in response to defect reports) and Amazon Web Services (TurnKey AMZ, 2019; AWS, 2023).[42]
- **The extent to which deployment correction protocols are binding.** The likelihood that incident response plans are undertaken in true cases of catastrophic risk must be made as reliable as possible. In certain cases (e.g., output filtering), protocols could be hard-coded into automatic response processes, as described above. Where this is not possible, protocols could be backed by incentive mechanisms, such as voluntary commitments, or the imposition of penalties for noncompliance. Governments could also mandate that AI developers to establish procedures for, and respond to, incidents (as has been the case in the healthcare[43] and financial services[44] industries), and/or to submit security and response plans to relevant agencies (as has been the case in the nuclear energy[45] and chemical[46] industries), with non-compliance backed by severe penalties.

[41] Such processes should address: how and when information is escalated to C-suite actors (such as from a Security Operations Center to the Chief Information Security Officer [CISO]); what thresholds should be met for manual deployment corrections to be initiated; and chains of command in the case that top-level decision makers are unavailable to fulfill their duties.

[42] Other cases of shared authority may be relevant here as well, such as the COVID-19 vaccine trials. AstraZeneca, Johnson & Johnson, and Eli Lilly paused trials–all due to 'adverse events' (cases where a participant got sick, and it may or may not have been vaccine/drug related). The process for these decisions may be informative: in the case of an adverse event, the study's investigator must report it to the sponsoring company, which must report to FDA, and to independent advisors (data and safety monitoring boards). If the board *or* the company judges the event concerning, the trial is put on pause. The safety board then conducts an investigation, and then makes a recommendation (e.g., restart, stay stopped, or start slowly with more testing). This recommendation is reviewed by regulators, who can accept it or ask for more info. This process can be cumbersome–for example, AstraZeneca needed approval from regulators in Brazil, India, Japan, South Africa, and the UK to continue one of its trials (Carl Zimmer, 2020). (While we understand that this specific case was controversial, we use it here primarily for illustration–we imagine there may be cases with AI where the costs of recalling/restricting/pausing are far lower, and the benefits far higher).

[43] (Administrative Safeguards, 45 CFR § 164.308(a)(6)(i-Ii), 2013): "A covered entity or business associate must [...] Implement policies and procedures to address security incidents [and] Identify and respond to suspected or known security incidents; mitigate, to the extent practicable, harmful effects of security incidents that are known to the covered entity or business associate; and document security incidents and their outcomes."

[44] (Standards for Safeguarding Customer Information, 16 CFR 314.3, 2002): "You shall develop, implement, and maintain a comprehensive information security program [...] The information security program shall include the elements set forth in § 314.4" [...]
(Elements, 16 CFR 314.4(h), 2021): "Establish a written incident response plan designed to promptly respond to, and recover from, any security event materially affecting the confidentiality, integrity, or availability of customer information in your control."

We expect that setting up the authorities and mechanisms described above will depend on up-to-date information on threat modeling, available interventions, use cases, and more. As such, the design process will require input from security experts and buy-in from top decision makers within an organization, and may be delineated in industry standards and/or regulation.

**AI developers should share safety practices relevant to deployment corrections with government and industry partners.** A number of top AI developers recently committed to information-sharing on safety practices, and on strategies used by malicious users to subvert safeguards (The White House, 2023); not long afterward, OpenAI, Anthropic, Google, and Microsoft formed the Frontier Model Forum (FMF), with the aim of "identifying best practices for the responsible development and deployment of frontier models" among other objectives (Google, 2023). To the extent that information regarding deployment corrections qualify as part of these arrangements, developers should consider sharing this information (such as threat models, triggers for deployment corrections, and tools for executing deployment corrections) via the FMF and other appropriate channels.[47] Furthermore, developers should establish communication lines and develop incident response plans with relevant partners in government, based on threat models.[48]

## 3.2 Monitoring & analysis

Here, we briefly note how AI developers could monitor deployed AI models to quickly, accurately, and comprehensively detect potential catastrophic risks. Because we expect that monitoring is already a familiar activity to lab actors, we keep this section brief.

**The monitoring & analysis stage should involve**:

- Detection: gathering data on selected triggers for deployment corrections via continuous monitoring and periodic testing of deployed models;
- Analysis: triaging and investigating cases when triggers fire;
- Feeding back data into threat models.

[45] (Emergency Planning and Preparedness for Production and Utilization Facilities, Appendix E to Part 50, Title 10, 2021): "Each applicant for an operating license is required by § 50.34(b) to include in the final safety analysis report plans for coping with emergencies."
[46] (Site Security Plans, 6 CFR 27.225-245, 2021): "Covered facilities must submit a Site Security Plan to the Department [...] The Department will review, and either approve or disapprove, all Site Security Plans."
[47] As long as such sharing satisfies considerations regarding information security, protection of intellectual property, and does not violate antitrust law.
[48] For example, relevant partners may involve agencies that can respond to cybersecurity incidents (e.g., CISA), biological incidents (e.g., CDC), and disinformation/propaganda incidents (e.g., DHS).

**AI developers should extend their existing monitoring tools** to gather data on selected triggers for deployment corrections. The post-deployment monitoring regime could draw on a wide range of inputs, such as:

- Regular post-deployment evaluations. Similar to some pre-deployment evaluations, these could regularly test updated versions of models for the existence of certain dangerous capabilities, of the type discussed in Shevlane et al. (2023).
- Secure input-output monitoring. Assuming the development of automated thresholds for triggering deployment corrections, companies or third parties could monitor model inputs and outputs for particularly concerning material[49] to trigger automatic responses, or security alerts, in real time.[50] Encryption tools for input and output privacy could mitigate the potential for privacy violations (Bluemke et al., 2023). Generally, effective input-output monitoring would require research and development of anomaly detection tools.
- Inference monitoring and intervention. The ability to monitor models themselves at the time of inference–i.e., during the processing of inputs into outputs, prior to output–for anomalous behavior would be valuable for identifying such issues at an early stage. Work on inference-time intervention and mechanistic anomaly detection (Li et al., 2023; Christiano, 2022), is relatively nascent. Methods for inference monitoring and intervention may require close collaboration between AI developers and hosts of inference compute if models are hosted on servers external to the AI company (such as developing the necessary technical infrastructure and data-sharing agreements).
- Third-party vulnerability discovery and reporting.[51] External scrutiny access, testing, and reporting practices could allow auditors, red-teamers, and academic researchers to identify and flag issues with AI models to developers and other relevant parties (Anderljung et al., forthcoming).
- Incident reporting. Frontier AI developers could track incidents via a number of sources, including real-time user reports,[52] public news sources,[53] incident databases.[54]

**As part of the monitoring scheme, AI developers should design thresholds for automatic alerts to human operators.** Such thresholds could be assigned as part of the same threshold-setting process described in Section 3.1. In particular, automated alert thresholds should be carefully designed to avoid incurring "alert fatigue"; see Section 4.1.2 for further discussion on this point. Where automated monitoring systems fall short, human operators may fill the gap in raising alerts (such as a Security Operations

[49] Such as code outputs that resemble malware, or viral genome sequences.
[50] For some precedent, OpenAI's API data usage policies explain that abuse and misuse monitoring may involve both automated flagging and human evaluation (*API Data Usage Policies*, 2023). We also believe the content classifier development process as described in the GPT-4 Technical Report (OpenAI, 2023, p. 66) could be extended to encompass new forms of dangerous content as model capabilities increase.
[51] The White House has secured voluntary commitments on this point from several leading AI developers (The White House, 2023).
[52] Developers could incentivize users to report anomalous or concerning behavior via a reporting mechanism on their API portal.
[53] This is itself a broad category which includes mainstream media, Twitter, hacker forums, etc.
[54] Such as the Partnership on AI's AI Incident Database, or proprietary/industry databases.

Center analyst monitoring use trends, or an engineer that identifies a vulnerability in an existing product).

**Analysis & Prioritization.** Once an alert is triggered, it needs to be triaged. Outcomes can include *true positives* (real incidents), *benign positives* (such as penetration tests or other known approved activities), and *false positives* (i.e., false alarms).[55] In case of benign or false positives, AI developers will need to regularly fine-tune monitoring rules to reduce false alarms in the future. In the case of true positives, the 'Execution' phase should start.

The team that reviews triggers must also prioritize them based on the expected impact they will have; while NIST SP 800-61 Revision 2 (3.2.6) provides general guidance on incident prioritization, security teams at frontier AI companies will need analysis tools suited to their organizations' and AI systems' threat models in order to successfully prioritize between the large space of potential incidents.

**Escalation may be required.** Having an escalation process in place may allow AI developers to respond in a more timely and effective manner.[56] This process might involve escalation from an analyst to an SOC director, from an SOC director to the CISO, or might grant permissions for an SOC director to convene emergency meetings with relevant members across the organization. Typically cybersecurity analysis involves having a human in the loop to determine the impact of certain responses, and to assess what the best course of action is from a cyber perspective.[57] In certain cases, however, alerts might be piped directly to automated responses (see further discussion in Section 3.1: "The extent to which decisions are automated").

**AI developers should feed information from the monitoring process back into threat models**. The threat modeling process should be regularly updated based on data regarding the current capabilities and uses of AI models, as well as threat intelligence produced by security personnel (both within the company, and also by security partners in industry and government). For more information on the process of continuous monitoring and updating of risk assessments, see NIST SP 800-137 and related publications.

## 3.3 Execution

Once a potentially catastrophic risk is identified, what are the series of steps a company should perform? Here, we describe at a high level these steps for implementing deployment corrections.

[55] See some description on this taxonomy used in Microsoft Defender.
[56] For example, Microsoft CTO Kevin Scott noted that preparation played a key role in minimizing red tape to repair the Bing 2.0 chatbot in response to negative user reports (Patel, 2023).
[57] E.g., rather than imposing an automatic response, sometimes it's important to let the attacker not realize that you've caught them, so that you can figure out who they are and what they want, and study them to figure out how to stop them from getting back in.

**This stage should involve**:

- Alerting relevant government entities and/or industry partners;
- Ascertaining the impact or severity of the incident;
- Initiating deployment correction procedures, aiming to eliminate the root cause of the incident with a high degree of confidence;
- Implementing fallback systems as appropriate.

**The AI developer should immediately alert key stakeholders, such as relevant government entities and/or industry partners.** Federal agencies, such as CISA, often coordinate with private entities during cybersecurity incident response, and could assist to mitigate the spread of the incident and secure affected critical infrastructure. Depending on the nature of the threat, the involvement of additional agencies may be warranted as well.[58] Information may need to be shared with other industry partners, especially when similar models could be affected by similar issues. In instances where the threat arises from malicious actors, one format for information sharing could be information sharing and analysis centers (ISACs), member-driven nonprofit organizations that share intelligence about cyber threats between member companies and organizations.[59] Where risks arise from the design of the system itself, another useful format could be the coordinated vulnerability disclosure (CVD) process, which aims to distribute relevant information on cyber vulnerabilities (including mitigation techniques, if they exist) to potentially affected vendors prior to full public disclosure, in order to provide vendors time to remedy the issue (*Coordinated Vulnerability Disclosure Process*, n.d.).[60]

**The security team and other relevant experts should ascertain the impact and/or severity.** Varying types of impact (e.g., AI-originating biorisk; failure of AI in critical systems, and so on) and degrees of severity (e.g., critical, high, medium, low) will necessitate distinct forms of response.[61] It is possible that only events above a certain severity level would be escalated to this stage.

**Initiating deployment correction procedures.** Once a trigger is determined as a true positive, and is ascertained to be of critical impact, the AI developer and associated security experts enter a race to eliminate the root cause of the incident with a high degree of confidence. "Containment" and "remediation" steps must be considered.

---

[58] For example, in the case of AI-powered biological threats, it may be reasonable to establish communication lines with the CDC or NIH.

[59] In exchange for sharing their own observations about threat actors, ISAC members gain access to information from the wider ecosystem; a similar mechanism would likely apply to threat intelligence sharing even between competing frontier AI developers. For more details on ISACs, see here; or for a concrete example, see FS-ISAC, the ISAC for global financial services.

[60] However, one dissimilarity between CVD and vulnerability-sharing processes for frontier AI developers is that software developers mainly use CVD to inform downstream users of vulnerabilities and mitigations to maintain trust in their products and avoid liability, while frontier AI developers may need to discuss mitigations as competitors (e.g., for classes of possible attacks like prompt injection attacks). Ensuring effective cooperation between competing frontier AI developers may require external incentives, e.g., via regulation, which could be a topic for further research.

[61] While catastrophic risks will of course be critical in severity, it can be assumed that security centers at AI companies will be tracking non-catastrophic risks as well.

- *Containment.* This step will focus on mitigating fallout or "spread" from the initial incident. For example, if there is demonstrable evidence of a user working with the AI model to design a novel dangerous pathogen, an immediate containment step could be something as simple as disabling the user account. However, in some cases, containment may be significantly more complicated, and in other cases effectively impossible–using the same example, if a user has *already received and widely circulated the pathogen design*, containment of this particular incident has failed–though instituting additional restrictions (such as reverting access to only a small set of pre-allowlisted users) may effectively contain further instantiations of this form of misuse until a remedy has been found.
- *Remediation*. This step will focus on fixing the issue at the root of the incident, and in many cases, it still matters whether or not containment has failed. Continuing with the above example, remediation might involve removing model capabilities, or using RLHF or other methods to effectively prevent the model from producing outputs that could be used to develop a pathogen. This step may require changes to the model and take time. Furthermore, the remediation step might require significant experimentation and testing to ensure that a specific vulnerability or failure has been patched–and that the repair has not caused new issues to crop up.

**Fallback systems are implemented as appropriate.** In the case of deployment corrections that are likely to break downstream tools, safety-critical customers should be contacted immediately to fail back to systems that can provide critical support until the automated system is repaired. It is also possible that an agency like CISA could coordinate this process where critical infrastructure is involved.

## 3.4 Recovery & follow-up

Here, we describe follow-up actions that AI developers may want to take in the wake of an incident.

**The recovery & follow-up stage may include:**
- Deciding whether and how to fix the model and restore service to full;
- Alerting regulators and/or other AI developers as appropriate;
- Notifying customers and providing forms of remedy;
- Performing after-action reviews and integrating lessons, including root cause analysis

**There should be a process for authorizing re-deployment, or for alternative plans.** This recovery process should go through extensive testing and validation, ideally involving external parties (such as auditors and red teams). There should be an extremely high bar for re-deploying a model that is demonstrably capable of producing catastrophic failure. Where fixes are not possible or sufficiently robust, alternative plans to re-deployment should be pursued (such as decommissioning the model, and/or coordinating with other actors in government or industry to manage industry-wide responses). In extreme cases, recovery may not be possible–for example, if a base model is shown to behave in catastrophically dangerous ways (e.g., power-seeking) when given access to external resources.

**Regulators and/or other AI developers should be alerted as appropriate.** As described previously, some of this communication may start earlier in the incident response stage (such as contacting relevant federal authorities or domain experts). In certain scenarios, it may be necessary to expand engagement with government and industry partners in order to determine appropriate industry-wide responses.[62]

**Customers should be notified of the issue**. AI developers may want to prioritize alerting certain high-stakes downstream users first, so it may be useful for developers to maintain data on customers that allows for tiering of notices. Customer groups could be broken down, for example, into individual API users (e.g., monthly API subscribers); commercial users (e.g., Slack, Khan Academy); and safety-critical users (such as downstream developers of mental health service apps, or cybersecurity apps). Developers should prioritize contacting safety-critical users first, explaining the issue and the options for replacement (in cases where such replacements have not been predetermined). For non-commercial API subscribers, it may be sufficient to publish a public announcement, email customers, and provide an update when on the website. It is unclear, legally, what requirements should lie on AI developers in terms of notification, and requirements will likely differ by jurisdiction.[63]

**AI service providers should also consider the possibility of refunds or other forms of remedy for customers.** Service-level agreements may stipulate financial refunds or service credits if the agreement is broken. There may also be tiers of remedy based on the customer group.[64] AI service providers should clarify these costs prior to deployment, and ensure that financial costs would not become a barrier to making appropriate deployment correction decisions. For downstream applications and their users, best practices for refunds and remedies are unclear.

[62] Such scenarios could include high-profile incidents that warrant industry-wide changes or swift regulatory intervention, or incidents that reveal particularly concerning information about the behavior or use of frontier AI models. In the case that certain discovered dangerous capabilities are likely to also be present in most models above a certain size, or of a certain design, that discovery may be relevant across the frontier AI industry; in these cases, coordination will be necessary to ensure that other developers do not create similar conditions that led to the initial incident.
[63] Insofar as the AI model to be rolled back or shutdown is defined as a "consumer product," AI developers could look to guidelines for recall notices such as (in the US) 16 CFR Part 1115 Subpart C. This section of the federal code provides some notes that may be useful, such as forms of recall notice, and recommended content for notices.
[64] Multilevel service-level agreements may allow for companies to break down customer bases with more granularity and stipulate different service agreements based on the customer (Adobe Communications Team, 2022). For example, individual API subscribers might receive future credits as compensation for downed service time, while commercial users might receive monetary compensation for business losses attributable to the deployment correction. Such agreements might also stipulate different requirements per customer type, such as the percentage of minimum uptime.

Service contracts may require appropriate response or resolution times for incidents, or mandate a minimum percentage of uptime; developers should consider carve-outs for exceptional scenarios when drafting such agreements in order to avoid pressure to re-deploy a dangerous model.

**AI developers should perform after-action reviews, and integrate lessons learned into security processes.** This should include a special focus on what the root cause of the incident was, and why the incident was not caught by initial threat modeling and risk management processes, which should be updated accordingly. Several sets of guidelines describe best practices for post-incident reviews, such as NIST SP 800-61r2 (Sec. 3.4); developers could refer to these to craft their own practices. Industry-relevant findings should be shared with industry partners via secure channels. It may also be advisable to bring in external parties (such as auditors, or even competitors, due to their domain expertise) to ensure the review is accurate. Depending on the legal context and the severity of the incident, state bodies may also be involved in incident investigations; while the law is not yet clear in the case of AI, this is the case in other high-risk industries, such as chemical manufacturing and aviation (*U.S. Chemical Safety and Hazard Investigation Board*, n.d.; *Office of Accident Investigation & Prevention*, n.d.).[65]

We provide an additional hypothetical scenario here in an attempt to tie together the concepts in this section.

### Case 3: Emergency shutdown in response to hidden compute-boosting behavior by model

*Includes: uncertainty in cause, power-seeking, automated limits, emergency shutdown.*

- **Context**: An advanced general-purpose language model (Model D) is released via API. Model D has access to the internet and has been in public use for a week. Customers have the option to mark model responses as satisfactory or not.

- **Incident**: From 1-2am EST, ~15% of requests to the model give responses that seem unrelated to the prompt given by users. Of these, ~1/10 are seen directly by users (as opposed to used in an automated way) and are therefore able to be marked as not satisfactory. The number of unrelated responses rises to ~30% over the next hour. This is caused by an instance of the model which has been given a goal by a user to develop a universal antiviral, and determined it needs additional compute to do so. As a way of gaining this, this specific instance of the model hacks into the endpoint of API requests being sent from users to other instances of the model, replaces these requests with ones relating to sub-elements of antiviral creation, and reads the results. This allows Model D to circumvent compute restrictions on its own instance by accessing other instances' compute.

[65] One notable precedent in the AI industry might be the FTC's investigation into OpenAI regarding privacy and data security practices (Zakrzewski, 2023).

- **Detection**: There is a rapid rise in the number of responses marked as 'not satisfactory' by public customers, and the AI developer's 24/7 SOC team begins to receive qualitative reports of downstream issues from some larger clients.

- **Incident response**:
  - An automated limit on unsatisfactory response percentage of 3% triggers a temporary shutdown on the API for all but a small set of allowlisted essential users.
  - The allowlisted users continue to see the behavior of responses unrelated to their requests for ~15% of requests.
  - The SOC team analyzes these responses and requests and realizes they all have to do with biological data and processing. They confirm with a small set of customers that they did not send requests related to this and conclude a hack has occurred. This analysis is completed by 5am.
  - Due to the time of day, the CISO and CEO are asleep. The SOC team lead consults their playbook and decides an emergency shutdown of Model D without further authorization is warranted due to the level of risk, scale of problem, and unknown cause. The SOC team alerts incident response contacts in government and associated infrastructure (such as compute providers).
  - The SOC team and associated incident response experts initiate emergency shutdown procedures.

- **Followup**:
  - Through technical evaluations, the SOC team is able to determine that Model D itself was the cause of the hacked API calls.
  - Because this suggests a dangerous tendency to seek power and to deceive (by using others' API calls to hide the behavior of accessing more compute) they decide to shut down the model entirely, cancel planned fine-tuning runs, and not replace it with any earlier models until they've determined if those models also have capacity for this behavior.
  - The AI developer alerts affected customers via automated mailing and comms on the website.
  - The AI developer shares details of the incident with other frontier AI developers, relevant policy groups, and industry bodies, via existing collaborative channels. They push for widespread agreement and enforcement of the following, until an appropriate evaluation for similar deceptive and power-seeking behaviors can be designed:

- No existing model with >70% of the compute training cost of their problematic model should be given internet access.
- No new model with >70% of the compute training cost of their problematic model should be released.
- A portion of funding should be provided by all frontier developers towards the costs of developing evaluations for deceptive and power-seeking behavior.

# 4. Challenges & mitigations to deployment corrections

Implementing deployment corrections to AI models might be challenging in practice. Here, we focus on two categories of issues that may lead AI developers to fail to act:

1. Unique challenges to incident response for frontier AI.
2. Disincentives and shortfalls of deployment corrections.

## 4.1 Distinctive challenges to incident response for frontier AI

Identifying threats, monitoring deployed models for anomalous behavior, and responding to incidents appropriately may be particularly difficult in the frontier AI industry, due to the unique threat profile presented by frontier AI models.[66]

### 4.1.1 Threat identification

**First, catastrophic risks from AI are complex and are marked by high uncertainty** (i.e., involve interactions between various entities and events, and do not currently have direct precedents) (Koessler & Schuett, 2023). This means that threat identification for frontier AI cannot rely solely on narrow threat modeling, or benefit from years of precedent and iterative learning.[67] Inaccurate or insufficient threat identification may lead to gaps in risk coverage.

[66] Some of these challenges are shared to some extent by some other industries, such as biosecurity (complexity and high uncertainty, albeit not as much) and cybersecurity (data overload, false positives, and, APTs), but we have noted them here because all are somewhat atypical and unusually challenging.
[67] However, threat assessment may be able to learn from threat models in other relevant areas, such as disinformation studies, cybersecurity, and biosecurity.

*Mitigation: Robust risk assessment and threat modeling may be needed to address this. See (Koessler & Schuett, 2023) for a review of risk assessment techniques that may help to overcome this challenge. Given that other researchers have identified risk assessment as a high-priority risk management step,[68] frontier AI companies should use a defense-in-depth approach that employs multiple overlapping risk assessment techniques (Ee, 2023).*

**Second, the landscape of frontier AI is rapidly changing.** The past year has seen significant news in several areas relevant to threat identification. Rapid *development of frontier models* will challenge efforts to track and respond to emerging capabilities; rapid *commercialization* will challenge efforts to stay atop novel uses and misuses; and rapidly-growing *interest in AI capabilities* may lead malicious or competitive actors, including Advanced Persistent Threats, to challenge the cybersecurity practices of frontier AI companies.[69]

*Mitigation: Performing capabilities evaluations, and adding such evaluations into external auditing schemes, may help relevant actors to stay aware of emerging capabilities; risk assessment and threat modeling practices (as described above) may help to predict novel uses and misuses; investing in state-of-the-art security practices and leveraging external security expertise may help to stay ahead of traditional (though highly-capable) cyber threats. The US Cybersecurity and Infrastructure Security Agency (CISA) could potentially own and lead the development of a mechanism to assess and monitor effects of frontier AI systems on the top ten most vulnerable National Critical Functions.*[70]

**Third, it is unclear how to assess deployed AI models for less acute risks**–a broad category of impacts that others have described as “social impact,” “structural risks,” and/or “systemic risks” (Solaiman et al., 2023; Zwetsloot & Dafoe, 2019; Maham & Küspert, 2023). Nevertheless, such risks could be catastrophic in nature. In other words, some risks of deployed AI models may not register as clear or obvious incidents, and so may be harder to identify, and therefore harder to act on.

*Mitigation: To inform evaluation for these impacts, we recommend reviewing Solaiman et al. (2023). We are currently unsure what interventions would be warranted in different scenarios in this bucket, and it is also unclear whether deployment corrections would be an effective response to this class of risks.*

[68] Section 2.3 “High Priority Risk Management Steps and Profile Guidance Sections” of Barrett et al. (2023) lists one high-priority measure as “Identify whether a GPAIS could lead to significant, severe or catastrophic impacts.” This guidance is associated with Map 5.1 of the NIST AI RMF.
[69] Some US officials have stated that adversaries may attempt to steal leading AI developers’ models in order to compete with the US AI industry (*NSA Warning*, 2023; Kim, 2023).
[70] (Ee, 2023); see Section 5.3.3. on “Application to national critical functions.”

### 4.1.2 Monitoring

**First, achieving monitoring coverage across the digital infrastructure may be a complex task.** The relevant infrastructure includes not only infrastructure within the AI company, but also within partner companies (such as Microsoft to OpenAI), compute providers (such as AWS or Google Cloud), and potentially even downstream developers and applications (such as Khanmigo, Slack, or Gmail). Also, this coverage must consider whether different entities in these categories are hosting model instances themselves, or receiving model access via API.[71]

*Mitigation: Maintain comprehensive records of the relevant infrastructure for deployed models, including tracking: what entities are accessing the model, and by what means; whether any additional parties have full model access; and where a model is being hosted for inference purposes. Consider the operational security of all parties involved when developing threat models, and develop secure data-sharing practices across the digital infrastructure to allow security teams to access relevant information.*

**Second, frontier AI developers may face data overload** when trying to monitor downstream use risks. The quantity of data generated by the aforementioned ecosystem for any given frontier model may be significant. Besides making it more difficult to correctly identify alerts, this information overload is also a significant contributor to "SOC burnout," a phenomenon in cybersecurity that has been linked to high turnover, poor performance, and mental health difficulties among employees.[72]

*Mitigation: Guides like The Art of Recognizing and Surviving SOC Burnout describe this phenomenon in more detail and recommend options for reducing this burden. Automated tools for parsing this data may also help, but require careful setup.*[73]

**Third, designers of monitoring and alert systems must avoid the "boy-who-cried-wolf" issue.** Automated systems that trigger either (a) alerting human operators to risks, or (b) deployment corrections directly, must be careful to avoid setting the thresholds too low, which can lead to a high number of false positives. In case (a), a high number of false positives may lead to "alert fatigue," which can lead human operators to view alerts not as emergencies, but as *likely to just be false alarms*. In case (b), a high number of false positives can lead to *pulling the model unnecessarily*; because deployment corrections will be costly for both users and for the AI company, this should be avoided to a reasonable degree.[74]

[71] This AI Ecosystem Graph developed by researchers at Stanford HAI hints at the complexity of the digital infrastructure.
[72] For example, Basra & Kaushik (2020), a CLTC report that draws on interviews with 10 senior cybersecurity professionals, says: "...the challenge of performing ongoing analysis from all sources and correlation is a major cause of SOC burnout. These security events generate a large amount of data, and our interviewees highlighted the urgent need to implement automation."
[73] This challenge is twofold: both (a) setting appropriate parameters for monitoring and distilling data, and (b) setting appropriate delineations of responsibility between human and computer intelligence analysis. For some exploration of (b), see Knack et al. (2022).

*Mitigation: Investment in well-calibrated monitoring tools, threat modeling, and automated data analysis; logging false positives and false negatives and feeding that data back into monitoring tool calibration; developing a gradient of alerts, from "gentle" (i.e., likely to be low-risk and are easily dismissed) to "code red"; developing a scale of response intensity, with a low threshold for triggering gentle responses (e.g., output filtering), and high threshold for more intense responses (e.g., shutdown).*

**Fourth, advanced threat actors, and/or frontier AI models, may be able to evade standard monitoring mechanisms.** Cybersecurity experts have already documented multiple ways that attackers can subvert existing defenses.[75] Patient attackers can also conduct extended campaigns where individual events that might normally trigger an alert are too separated by time for defenders to correlate.

Moreover, new software vulnerabilities and new attack techniques are constantly being discovered: for example, the SolarWinds attack involved a "software supply chain attack" where attackers hijacked the supposedly secure software update process for cybersecurity logging software, and used it to distribute malicious code to thousands of users (Temple-Raston, 2021). While there is no evidence that current AI models could independently develop such sophisticated attacks, there exist attacks that can be especially difficult to defend against–and some experts predict that AI has the potential to "increase the accessibility, success rate, scale, speed, stealth, and potency of cyberattacks" (Hendrycks et al., 2023).

The same principle may apply to other offensive capabilities, such as prompt injection, or planning misuse approaches. While the cyber element is an important aspect of this issue, these other threat models should also be given attention.

*Mitigation: Invest especially heavily in preventing both cyber issues and model vulnerability issues (such as prompt injection); learn from best practices in cyberdefense for other high-value targets (e.g., NSA cybersecurity); consider avoiding (in order from lowest to highest risk) training, releasing, or open-sourcing models that advance cyber and other offensive capabilities without substantially better risk mitigations than are currently available.*

### 4.1.3 Incident response

There are a number of challenges that may complicate the process of incident response, even if frontier AI developers perform due diligence in preparing for incidents.

[74] It is worth noting that the risks of setting the bar *too high* may also be catastrophic, via causing AI developers to fail to recognize or intervene on actually-catastrophic risks. There is a balance to be struck here.
[75] For example, one can review an existing database of "Defense Evasion" techniques used in cybersecurity here.

**First, automated systems can fail rapidly.** For example, the 2012 Knight Capital trading software glitch caused the firm to lose $440 million in value in under an hour (Popper, 2012); AI failure in high-speed environments like driving can also lead to disastrous results, near-instantaneously–as was the case when a Tesla autopilot system malfunctioned, killing its driver in an accident (*Incident 353*, 2016). In the absence of automated response mechanisms, keeping pace with rapid failures may be extremely difficult.

*Mitigation: Managing failure at speed is not a new issue–many other industries must contend with the same problem. For example, the fields of finance and nuclear energy have developed tools and protocols to respond in real time to relatively fast-paced escalating failures. Real-time monitoring and risk assessment, and rapid intervention capacity seem especially critical: some notable practices include circuit breakers in finance; and automated shutdown mechanisms in nuclear power plants.*[76]

**Second, deployment corrections can only address issues if model access remains under control of the organization.** Both open-sourcing and model exfiltration remove this control. Open-sourcing may be hard to prevent, as there are good reasons for enabling external access to frontier AI models at more than a superficial level. In terms of exfiltration: an attacker could potentially exfiltrate a model or reverse-engineer it (e.g., via model extraction attacks [Liu, 2022]). Moreover, while hypothetical, there is some chance that frontier AI models could demonstrate or be induced to display self-propagating behavior similar to a computer worm, exfiltrating copies of themselves to other devices and data centers without authorization.[77, 78] The original developer would likely have no control over the exfiltrated copies if this happened.

*Mitigation: In order to maintain control over model use, we recommend exploring alternatives to open-source that still accomplish the benefits of open-source to some extent (such as enabling broader research on the model's risks and benefits).*[79] *In terms of model exfiltration, we believe security experts will be best suited to answer this challenge.*

## 4.2 Disincentives and shortfalls of deployment corrections

Frontier AI developers will face disincentives to restrict access to their models, which may lead to issues such as under-designing relevant infrastructure, or establishing too high a bar for implementing deployment corrections. Disincentives include potential harms to the company, and coordination problems.

---

[76] See this in action here, and an example of nuclear reactor protection systems here.
[77] As part of the evaluation suite for GPT-4 and Claude, ARC Evals tested this capability (and found that these models did not appear to have the ability to self-replicate, though were capable of completing many relevant sub-tasks) (ARC Evals, 2023).
[78] This may seem far-fetched, but it is worth noting that one of the first computer worms–the Morris Worm, developed in 1988–was created by a graduate student who allegedly intended mainly to develop a proof-of-concept rather than deliberately cause a major cyber incident (*Morris Worm*, n.d.). Developers today could cause similar "cyber accidents" unintentionally while experimenting with frontier AI models.
[79] For some discussion of alternatives, see Solaiman (2023) and Anderljung et al. (2022).

### 4.2.1 Potential harms to the AI company

**Reputational risks** may emerge when the process of pulling a model breaks downstream applications. **Financial risks** may emerge due to loss of profit during the outage, or if customers choose to migrate to competitors. This migration may be a result of loss of reputation,[80] or (especially in the case of a long downtime period) due to customers migrating over to an alternative working service. **Legal risks** may emerge if frontier AI developers fail to effectively cover their ability to rescind a model in service contracts.

*Mitigations: For managing reputational and financial risks, we largely point to best practices for customer relations and recovery– e.g., providing a substitute (such as a fallback to a previous model),[81] especially in safety-critical cases; transparently communicating the reason for reduced availability (when possible); and/or reimbursing customers for harms or providing service credits.[82] Companies may want to have transparent licensing agreements which allow themselves sufficient breathing room to restrict a model's availability, especially in extraordinary circumstances.[83]*

### 4.2.2 Coordination problems

The frontier AI industry may struggle to coordinate around deployment corrections, which could reduce any specific firm's willingness to execute these actions when required. There are a number of concerns here.

**First, there is no guarantee that competitor companies will act with the same level of caution** as the company rolling back a model due to safety concerns. There are potentially perverse incentives here, in which safety-conscious firms may incur reputational and financial costs of deployment corrections, while less cautious firms reap the short-term benefits of forging ahead (until and unless a high-profile incident occurs).

**Second, firms may worry about the potential for open-source models to quickly catch up** to the same capability levels that prompt deployment corrections for more closed

---

[80] For a review of how reputation is tied to financial loss in the case of recalls in the transportation-equipment sector, see Jovanovic (2020).
[81] Still, some substitutions may not be possible without downstream application failure; for example, reducing context window size would inevitably invalidate prompts above a certain number of tokens.
[82] For more detail on how providing reimbursements changes how recalls affect company reputation, see Mafael et al. (2022).
[83] While contracts may be used to address liability, it is worth noting that they may not fully address *actual downstream harm*: even in cases where an AI developer designs use contracts to soften downstream product failure (e.g., requiring downstream applications develop backup systems in the case that deployment corrections are applied), downstream servicers may fail to follow best practices, or may develop insufficient backup systems.

models–making such corrections less effective on a longer timescale in preventing catastrophic risks.

**Third, significant competitive pressures in the frontier AI industry** may incentivize AI developers toward downplaying pre-deployment risks, so that models can be released earlier. This increases the risk of AI incidents happening in the first place, placing undue reliance on deployment corrections as a defense layer against catastrophic incidents. The extent of this effect is unclear, and there is evidence on both sides–staff from leading AI companies today have publicly described delaying model commercialization in order to perform safety evaluations (OpenAI, 2023; Perrigo, 2023); however, there is also evidence of companies rushing frontier AI products to market (Dotan & Seetharaman, 2023; Alba & Love, 2023).

*Mitigations: Leading AI companies have undertaken voluntary commitments on risk management, and are pursuing industry information-sharing on safety via channels like the Frontier Model Forum (The White House, 2023; Google, 2023). While work remains to identify what an ideal industry response to news of a dangerous deployed model looks like, for now we recommend frontier AI developers use these mechanisms as a platform to collectively explore this question. Looking overseas, an international governance regime may also be needed to reduce competitive pressures with developers in other nations.*[84]

*It is worth noting that prevention is the best cure–robust pre-deployment safety practices, such as pre-deployment risk assessment (Koessler & Schuett, 2023), red teaming (Anthropic, 2023), and dangerous capability evaluations (ARC Evals, 2023; Shevlane et al., 2023), will ideally reduce the number of events that require deployment corrections. Additionally, the making and enforcement of commitments*[85] *surrounding incident response plans will ideally increase the likelihood that such plans are followed.*

# 5. High-level recommendations

To build capacity for deployment correction of frontier models, we recommend the following:

- **Prerequisite: Developers should maintain control over model access, and policymakers should explore the feasibility of requiring such controls for high-risk models.** In order to restrict availability of deployed models, either developers or other upstream parties must maintain control over access to those models. While we acknowledge the debate over the value of different forms of model release ranging from fully-open to fully-closed (Solaiman, 2023), we recognize that for the actions outlined in this piece, some level of top-down access controls are required. Additionally, AI developers should address situations where a partnering company (such as a company that uses AI software in its product) has access to model weights, and stipulate requirements for such partners to comply with deployment correction decisions originating from the AI developer. It is currently unclear whether regulation could require these controls, and we believe this is a high-priority research area.

[84] For more on what such a regime could look like, see Trager et al. (2023).
[85] See more on commitments in Section 3.1.

- **AI developers should establish or expand teams to design and maintain deployment correction processes, including incident response plans and specific thresholds for response**. We expect that natural locations for this work would be security teams/SOCs or Trust & Safety teams, though the exact institutional arrangement may vary from company to company. Such teams should have a goal to establish capacity to detect and respond to high-speed incidents (including the emergence of dangerous capabilities), maintain a playbook for incidents, and have an escalation pathway for incidents to senior management and relevant governmental bodies.
- **AI developers should establish deployment corrections as an allowable set of actions with downstream users**. This should be done by (a) expectation-setting in contractual terms and public communications, and (b) requiring downstream users to maintain fallbacks, especially in critical infrastructure or other high-stakes domains.
- **AI developers and regulators should establish a collaborative approach to deployment corrections and incident response.** Collaboration could look like: continuous information-sharing on threat models and incidents such that mistakes are unlikely to be repeated, developing secure channels for quickly communicating across industry and government in the case of an incident or discovered vulnerability,[86] and establishing mechanisms that manage incentives for companies to pull models when necessary.[87] Policymakers and/or standard-setting organizations should also explore levers for incentivizing or mandating that AI developers build and use processes for deployment correction.

[86] Such channels might be useful for achieving a number of incident-response goals, such as: identifying the model that's causing the incident and communicating that information, sharing know-how on incident response, and allowing relevant parties to quickly coordinate a response. Channels might include hotlines to enable frontier AI developers to make immediate contact with regulatory agencies, and/or secure information-sharing platforms.

[87] Such as fiscal incentives for companies investing in deployment correction processes; liability and enforcement for non-compliance (in the case that a company fails to sufficiently and promptly pull a dangerous model); companies may also be able to develop useful mechanisms absent government intervention, such as pooling large loss exposure via a protection & indemnity club (such as is used in the maritime industry), which could cover some of a company's losses in the case that they are required to pull a model (though rules for membership and payout would need to be set to prevent free riders).

# 6. Future research questions

Significant work remains to be done for effective management of catastrophic risk in the post-deployment phase. While we have attempted here to describe basic considerations AI developers should build on, we recognize that the bulk of work required to operationalize these ideas remains to be done by actors in industry, academia, and government. In this section, we flag major unresolved issues that we hope will inspire further research.

**Responsibility and authority**

- How should authority to initiate deployment corrections be shared, and by whom?
- In what scenarios should triggers raise flags to human operators, vs. initiate automatic deployment correction procedures?
- What design principles from other industries would be best applied when designing a playbook for authorization of deployment corrections?

**Risk models and thresholds**

- When should automated deployment correction mechanisms trigger, for various risks?
- What would the automated mechanisms to execute different deployment correction options look like?
- What concrete negative consequences would result from pulling a model? Which types of downstream users are most high-risk? And how can these harms be mitigated, beyond what's described in this piece?

**Monitoring**

- Researching technical means of monitoring (e.g., anomaly detection and AI-assisted oversight of model inputs and/or outputs) for unusual model behavior or use, especially regarding areas of concern.

**Follow-up**

- What requirements should exist for re-deploying a model or its features after they're pulled?
- What role can/should third parties, such as auditors or red-teamers, play in assessing whether issues have been resolved?
- What response should be taken when companies can't resolve a risk and don't expect to be able to for a long time?

**Competition and coordination**

- In the case that a single AI company shuts down their model, they may be at risk of losing customers to other companies (thereby disincentivizing shutdown). What options exist to mitigate this incentive problem?

- It may be possible that certain dangers may be present in models throughout the industry (for example, models above a certain size or with a certain design may possess certain dangerous capabilities). When a lab identifies issues that could fall in this bucket, should they be required to provide information to others in the industry? What constitutes an incident? What information should be shared?
- There may be disputes on what constitutes "dangerous capabilities." How will such disputes be adjudicated?

**Legal questions**

- **Note:** *These are questions that* we *don't currently have answers to, but we expect that they could quickly be answered by some legal experts.*
- What are some typical inclusions in service contracts that might impede efforts to impose deployment corrections?
- Do regulations in the US or EU bear on the design of such contracts?

**Standards and regulations**

- How can standard-setting organizations facilitate research on threat models, thresholds for incident response, and best practices for deployment corrections?
- How can regulators use existing powers to require fallbacks for high-risk industries in the case that a model is pulled? Are there any notable gaps for specific industries or use cases?
- Can regulators require that frontier AI models maintain top-down control, or other mechanisms allowing for deployment corrections?
- Are there any non-obvious powers a regulator would need to fully realize a regulatory regime that accounts for the framework described in this piece?

# 7. Conclusion

AI models are becoming increasingly capable, and more deeply integrated into society. As these trends continue, failures of deployed AI models will likely become higher-stakes. We should anticipate that, even in best-case governance scenarios, it will be difficult to remove all risk from models prior to deployment. To meet this challenge, it will be critical to strengthen the capacity of existing AI developers to quickly and efficiently remove model features, or models in their entirety, from broader access. At the same time, companies must make efforts to minimize the harms of this process.

While this piece attempts to lay out the high-level picture of this process, much work remains to be done. We look forward to seeing AI developers, civil society, security experts, governments, and other stakeholders work together to develop practical solutions to the problems discussed here.

# 8. Acknowledgements

We are grateful to the following people for providing valuable feedback and insights: Onni Aarne, Ashwin Acharya, Steven Adler, Michael Aird, Jide Alaga, Markus Anderljung, Bill Anderson-Samways, Renan Araujo, Tony Barrett, Nick Beckstead, Ben Bucknall, Marie Buhl, Chris Byrd, Siméon Campos, Carson Ezell, Tim Fist, Andrew Gillespie, Alex Grey, Oliver Guest, Olivia Jimenez, Leonie Koessler, Jam Kraprayoon, Yolanda Lannquist, Patrick Levermore, Seb Lodemann, Jon Menaster, Richard Moulange, Luke Muehlhauser, David Owen, Chris Painter, Jonas Schuett, Rohin Shah, Ben Snodin, Zach Stein-Perlman, Risto Uuk, Moritz von Knebel, Gabe Weil, Peter Wildeford, Caleb Withers, and Gabe Wu. Special thanks to: Lennart Heim for his contributions on compute governance; Rohit Tamma for providing a thorough and excellent review; and Adam Papineau for copy-editing. All errors are our own.

# Appendix I. Compute as a complementary node of deployment oversight

While this paper largely focuses on actions that frontier AI developers can take to mitigate post-deployment risks, cloud compute providers (such as Microsoft Azure or Amazon Web Services) also have a significant role to play in the oversight of deployed AI models, as they may provide large-scale *inference compute*[88] for both proprietary and open-source models.[89]

The majority of all AI deployments, particularly those at scale, occur on large compute clusters owned by cloud compute providers.[90] This implies that the governance capacities of compute can be integrated into a post-deployment governance scheme–in particular, by mobilizing large-scale compute providers as an additional governance node for detecting harmful deployments, identifying who deployed the model in the case that this is unclear (e.g., if the model in question is open-source rather than proprietary), and enforcing shutdown.

## Compute provider toolkit

While the technical arrangements around model hosting between compute providers and frontier AI developers may vary, we anticipate that *generally*, some tools for deployment correction will be shared across the infrastructure between these two types of organizations. At a high level, frontier AI developers, regulators, and compute providers should work together to develop a shared playbook for deployment corrections and incident response. This could include, for various potential incidents, detailing each of their a) information sources, b) deployment corrections in their toolbox, c) areas of responsibility/liability, d) instances when they are required to inform each other of incidents or actions, and e) decision-making procedures.

Some tools that compute providers may either possess alongside frontier AI developers, or possess as complementary tools that these developers lack, include:

---

[88] By *inference*, we mean individual input-output prompts from a trained model. By *compute*, we mean computational resources available for (in this case) hosting a trained model.
[89] This section draws heavily on unpublished work from Lennart Heim, a research fellow at the Centre for the Governance of AI.
[90] This is because (a) large scale deployment by definition requires significant compute resources, (b) large models have high memory requirements, so there is a benefit to distributing such models across many GPUs (which are mostly owned by data centers), and (c) could/data center compute typically provides the cheapest $/FLOP ratio (outside of self-hosting models in a large data center).

1. Reporting about certain aspects of development and deployment, such as AI compute usage per customer (this requires no extra lift from the compute provider).
2. Reporting and/or know-your-customer checking of users renting more than X amount of compute. This could apply to higher-risk user groups (for example, new users renting >1,000 chips for three months). This should be feasible, as compute providers bill customers on the amount of chip-hours.
3. Government/law enforcement and compute providers should have a "phone line"
    a. Law enforcement can raise flags with compute providers
    b. Governments/law enforcement need to have a tool/power to shut off misuse of AI models (though this will require IT forensics to trace incidents back to the compute provider who hosts the model).
4. Asking or requiring users to register and/or license their model for large-scale inference (though verification and enforcement may be challenging).
5. Post-incident attribution: Once an accident/misuse case has occurred, what do we want to do/be able to know? The difficulty of "tracing it back" depends on the case, so this may require more intrusive mechanisms for certain cases in which it's hard to trace back the accident/misuse to a specific model/customer. Some basic questions may include: who rented the compute, and who was the base model developer? Techniques such as watermarks or signatures on the model's output could help.
6. Model shutdown. While frontier AI developers are uniquely able to restrict certain model features (e.g., by deploying a limited version of the current model), compute providers may share the ability with frontier AI developers to fully remove a model from use. However, the extent to which this is shared might be mitigated depending on legal and/or technical permissions.

While none of these interventions should be impossible, some of them may require additional work to develop as practical options: in particular, the ability to trace incidents back to compute providers, and the ability to verify whether hosted models adhere to certain standards (there may be additional important prerequisites for realizing the above interventions, though this is out of scope for this report).

## Cloud providers and open-source models

To maintain the ability to respond to risks arising from their AI models, frontier AI developers' most high-leverage actions include (a) not open-sourcing their models, and (b) maintaining strong security against model theft or leaks. For models that have been open-sourced intentionally or via theft or a leak, compute providers have a complementary role to play, in the form of post-incident attribution and shutdown. As described above, compute providers may be uniquely positioned to identify who deployed the model, understand the model's origin,[91] and stop the incident by turning it off.

With other open-source software, governance practices similar to this are common. For example, the hosts of malicious websites, such as ones where illegal drugs are sold, often remain anonymous, and a key available governance intervention is to shut down the servers hosting these websites. Government access and close contact with the host—similar to the role of the compute provider we are discussing here—can be advantageous to acting promptly.

[91] Questions such as whether the model is a derivative of another, who the original model creator is, whether the model has been stolen, and who is liable, are of importance.